\documentclass[amsmath,amssymb,pra,showpacs,twocolumn,superscriptaddress]{revtex4}

\usepackage{graphicx}
\usepackage{times}

\newcommand{\bra}[1]{\langle #1|}
\newcommand{\ket}[1]{| #1 \rangle}
\newcommand{\op}[1]{\hat{#1}}
\newcommand{\cre}[1]{\hat{#1}^\dagger}
\newcommand{\des}[1]{\hat{#1}}

\begin{document}

\title{Cooling in the single-photon strong-coupling regime of cavity optomechanics}

\author{A.~Nunnenkamp}
\affiliation{Department of Physics, University of Basel, Klingelbergstrasse 82, CH-4056 Basel, Switzerland}
\author{K.~B\o rkje}
\affiliation{Departments of Physics and Applied Physics, Yale University, New Haven, Connecticut 06520, USA}
\author{S.~M.~Girvin}
\affiliation{Departments of Physics and Applied Physics, Yale University, New Haven, Connecticut 06520, USA}

\date{\today}

\begin{abstract}
In this paper we discuss how red-sideband cooling is modified in the single-photon strong-coupling regime of cavity optomechanics where the radiation pressure of a single photon displaces the mechanical oscillator by more than its zero-point uncertainty. Using Fermi's Golden rule we calculate the transition rates induced by the optical drive without linearizing the optomechanical interaction. In the resolved-sideband limit we find multiple-phonon cooling resonances for strong single-photon coupling that lead to non-thermal steady states including the possibility of phonon anti-bunching. Our study generalizes the standard linear cooling theory.
\end{abstract}



\maketitle

\emph{Introduction.} In optomechanical systems mechanical degrees of freedom are coupled to modes of the electromagnetic field inside an optical or microwave resonator \cite{Kippenberg2008, Marquardt2009}. Possible applications include ultra-sensitive sensing of masses, forces and electromagnetic fields \cite{Rugar2004}, transducing quantum information between different parts of quantum networks \cite{Rabl2010}, and exploring decoherence at larger mass and length scales \cite{Marshall2003}.

For these applications it is very important to minimize the influence of thermal fluctuations. This is why a large part of current experimental efforts is directed at cooling the mechanical degrees of freedom. Recently, the quantum ground state of mechanical motion was achieved for mesoscopic oscillators \cite{OConnell2010, Teufel2011, Chen2011} and the zero-point motion was detected by observing an asymmetry between phonon absorption and emission rates, originating from the fact that an oscillator in the quantum ground state cannot emit but only absorb energy \cite{Safavi2011, Brahms2012}.

In most optomechanical setups the position of the mechanical oscillator linearly modulates the cavity frequency. While the optomechanical coupling on the single-photon level is usually much smaller than the cavity linewidth, coupling between the fluctuations of the cavity field and the mechanical position can be made appreciable using a strong optical drive. For this situation a quantum theory of red-sideband cooling has been developed in Refs.~\cite{Wilson-Rae2007, Marquardt2007}.

Several experiments \cite{Weis2010, Teufel2011, Safavi2011, Brooks2011} are currently approaching the regime where the presence of a single photon displaces the mechanical oscillator by more than its zero-point uncertainty. Going beyond early work \cite{Mancini1997, Bose1997} novel effects in this regime have recently been predicted, including mechanically-induced cavity resonances \cite{Nunnenkamp2011, Rabl2011}, multiple mechanical sidebands \cite{Nunnenkamp2011}, photon anti-bunching \cite{Rabl2011}, non-Gaussian \cite{Ludwig2008, Nunnenkamp2011} or non-classical \cite{Qian2011} mechanical steady-states, and scattering \cite{Liao2012} of and interferometry \cite{Hong2011} with single photons. However, cooling of the mechanical oscillator in the regime of nonlinear strong coupling has not been discussed in the literature.

In this article we study how the weak-coupling cooling theory is modified in the single-photon strong-coupling regime. Using Fermi's Golden rule we calculate the transition rates caused by the coupling to the optical field without linearizing the optomechanical interaction. In the resolved-sideband limit we find cooling resonances if the cavity is driven on one of the several mechanical sidebands. In contrast to the weak-coupling regime the phonon transition rates do not obey detailed balance. We find steady states with non-thermal phonon number statistics including phonon anti-bunching.

Our study generalizes the standard theory of red-sideband cooling \cite{Wilson-Rae2007, Marquardt2007} to the regime of strong optomechanical coupling. In the literature nonlinear cooling has been discussed in the context of trapped ions outside the Lamb-Dicke regime \cite{Matos1994} and of optomechancial systems where the cavity is coupled to the position squared of the oscillator \cite{Nunnenkamp2010}.

\emph{Model.} We consider the standard model of optomechanical systems where the position of a mechanical oscillator, $\hat{x} = x_\mathrm{ZPF}(\des{b} + \cre{b})$, is parametrically coupled to an optical cavity mode $\des{a}$. The Hamiltonian reads
\begin{equation}
\hat{H}_0 = \hbar \omega_R \cre{a} \des{a} + \hbar \omega_M \cre{b} \des{b} + \hbar g \cre{a} \des{a} (\des{b} + \cre{b})
\label{Ham1}
\end{equation}
where $\omega_R$ is the resonator frequency, $\omega_M$ the mechanical frequency, and $g=\omega_R' x_\mathrm{ZPF}$ is the optomechanical coupling. $x_\mathrm{ZPF} = \sqrt{\hbar/(2M\omega_M)}$ is the zero-point uncertainty, $M$ the mass of the mechanical oscillator, and $\omega_R' = \frac{\partial \omega_R}{\partial x}$ the derivative of the resonator energy with respect to the oscillator position $x$. $\des{a}$ and $\des{b}$ are bosonic annihilation operators for the cavity mode and the mechanical oscillator, respectively.

In order to include drive and decay we use standard input-output theory \cite{GirvinRMP}. In a frame rotating at the frequency of the optical drive, the nonlinear quantum Langevin equations read
\begin{align}
\dot{\des{a}} &= + i \Delta \des{a} - \frac{\kappa}{2} \des{a} - i g \left(\cre{b} + \des{b} \right) \des{a} + \sqrt{\kappa} \, \des{a}_\mathrm{in} \label{motion1} \\
\dot{\des{b}} &= -i \omega_M \des{b} - \frac{\gamma}{2} \des{b} - i g \cre{a} \des{a} + \sqrt{\gamma} \, \des{b}_\mathrm{in}. \label{motion2}
\end{align}
where $\Delta = \omega_L - \omega_R$ is the detuning between laser $\omega_L$ and resonator frequency $\omega_R$, and $\gamma$ and $\kappa$ are the mechanical and cavity damping rates. The cavity input $\des{a}_\mathrm{in} = \bar{a}_\mathrm{in} + \des{\xi}$ is the sum of a coherent amplitude $\bar{a}_\mathrm{in}$ and a vacuum noise operator $\des{\xi}$ satisfying $\langle \des{\xi}(t) \cre{\xi}(t') \rangle = \delta(t-t')$ and $\langle \cre{\xi}(t) \des{\xi}(t') \rangle = 0$. Finally, we assume that the mechanical bath is Markovian and has a temperature $T$, i.e.~$\langle \des{b}_\mathrm{in}(t) \cre{b}_\mathrm{in}(t') \rangle = (n_\mathrm{th}+1) \delta(t-t')$ and $\langle \cre{b}_\mathrm{in}(t) \des{b}_\mathrm{in}(t') \rangle = n_\mathrm{th} \delta(t-t')$ with $n_\mathrm{th}^{-1} = e^{\hbar \omega_M/k_B T}-1$.

\begin{figure}
\centering
\includegraphics[width=0.49\columnwidth]{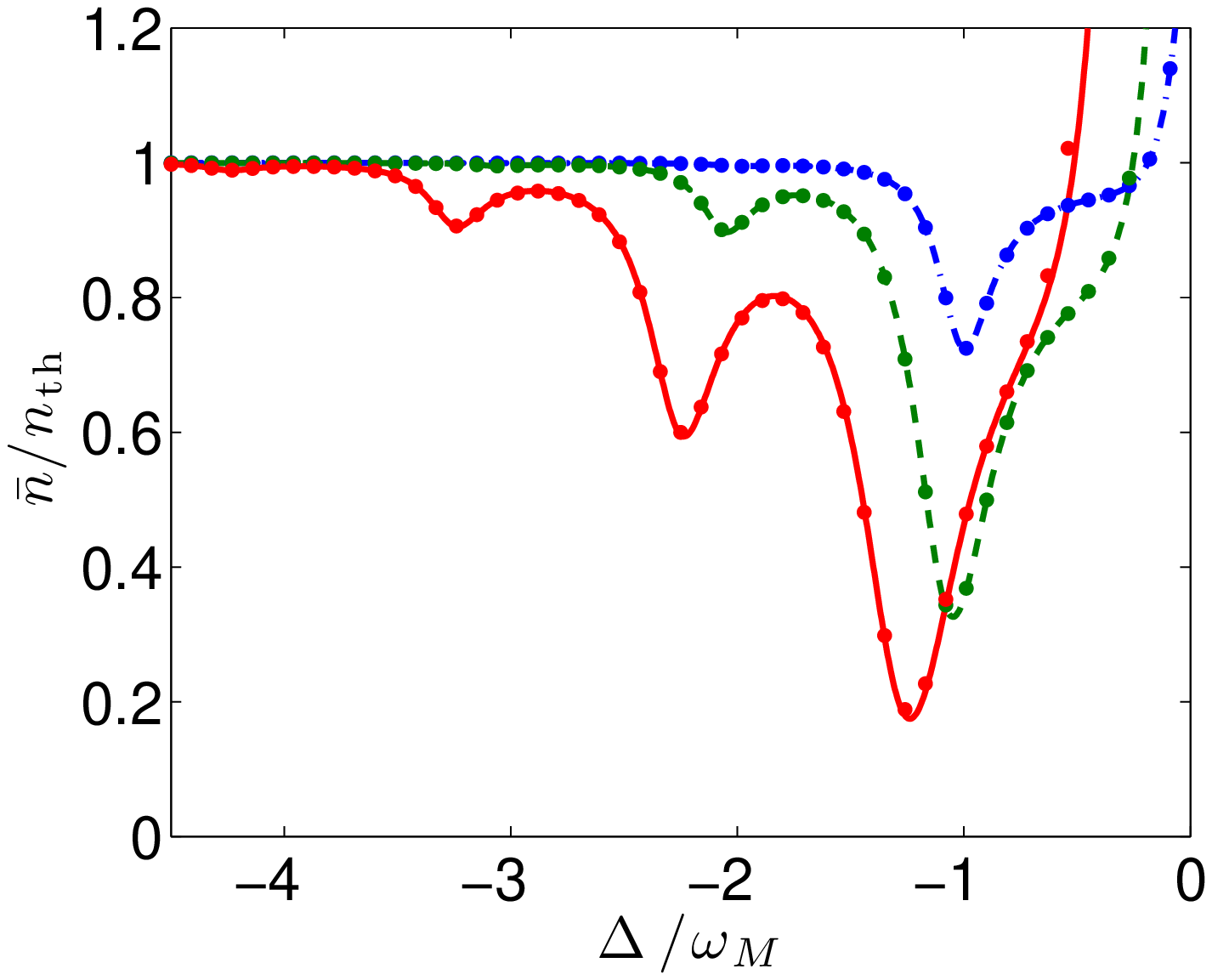}
\includegraphics[width=0.49\columnwidth]{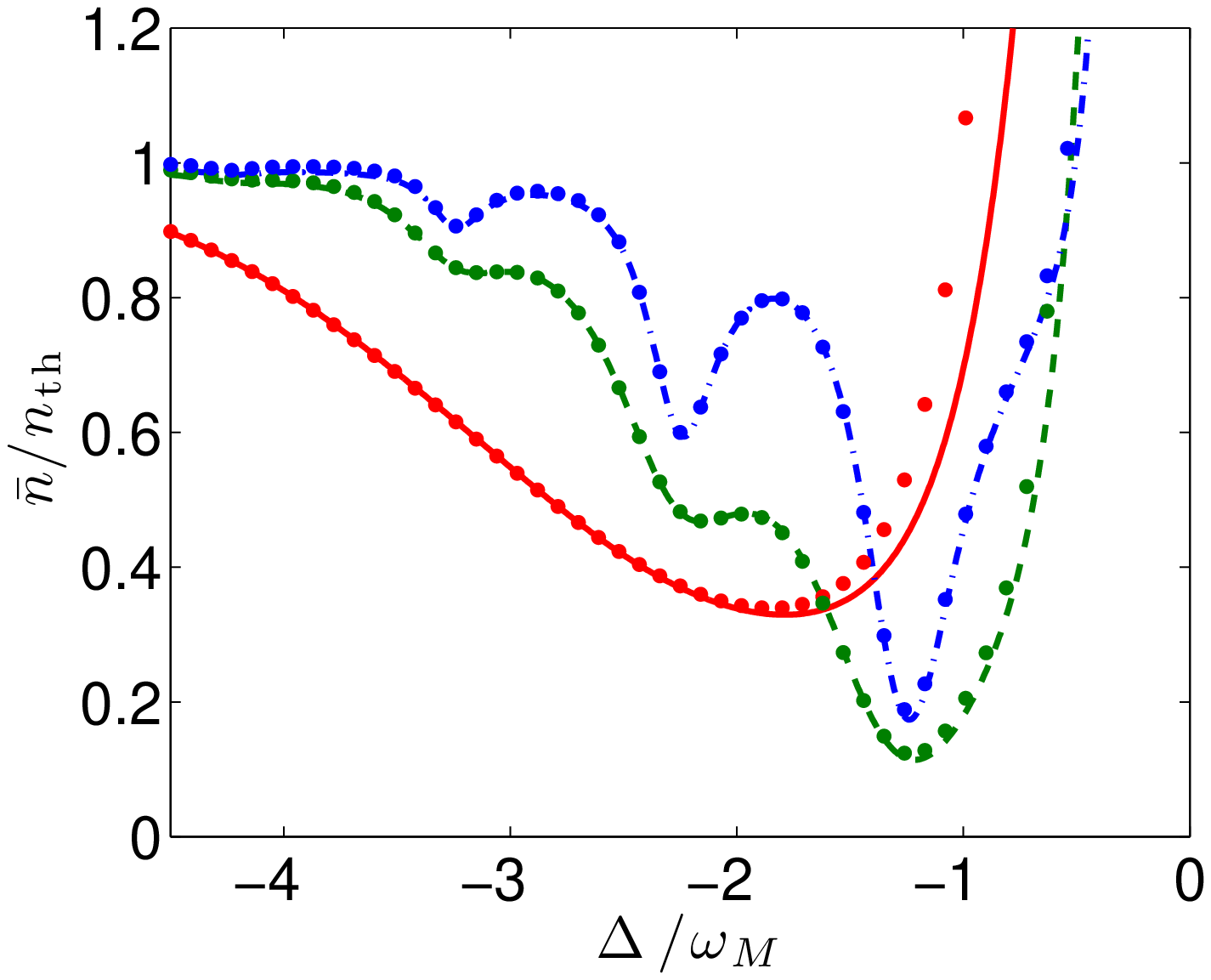}
\caption{(Color online) \emph{Multiple cooling resonances.} Steady-state phonon number $\bar{n} = \langle \cre{b} \des{b} \rangle$ (a) for $g/\omega_M = 0.5$ (red solid), $g/\omega_M = 0.25$ (green dashed) and $g/\omega_M = 0.1$ (blue dash-dotted) as well as $\omega_M/\kappa = 4$ and (b) for $\omega_M/\kappa = 0.5$ (red solid), $\omega_M/\kappa = 2$ (green dashed) and $\omega_M/\kappa = 4$ (blue dash-dotted) as well as $g/\omega_M = 0.5$. Results from the set of rate equations (\ref{rateeqn}) are shown as lines and those from the quantum master equation (\ref{master}) as dots. The other parameters are $\omega_M/\gamma = 1000$, $n_\mathrm{th} = 1$, and $\Omega/\kappa = 0.2$.}
\label{fig:cooling}
\end{figure}

\emph{Multiple cooling resonances.} We are interested in the influence of a weakly driven, but strongly coupled optical field on the mechanical oscillator. We calculate the mechanical cooling and amplification rates induced by the optical drive using Fermi's Golden rule and write down a set of rate equations for the mechanical oscillator
\begin{align}
\dot{P}_n &= - \gamma n_\mathrm{th} (n+1) P_n - \gamma (n_\mathrm{th}+1) n P_n \nonumber \\&
+ \gamma n_\mathrm{th} n P_{n-1} + \gamma ( n_\mathrm{th}+1) (n+1) P_{n+1} \nonumber \\ &
- \sum_{m\not= n} \Gamma_{n \rightarrow m} P_n + \sum_{m\not= n}  \Gamma_{m \rightarrow n} P_m
\label{rateeqn}
\end{align}
where $P_n$ is the probability of the oscillator to be in the state with $n$ phonons. The terms in the first two lines are due to the coupling of the mechanical oscillator to its thermal bath with rate $\gamma$ and thermal phonon number $n_\mathrm{th}$. The sums in the last line are the terms caused by the coupling to the cavity field.

In the frame rotating at the drive frequency $\omega_L$ the drive is described by the Hamiltonian $\hat{H}'_1 = \hbar \Omega (\hat{a} + \hat{a}^\dagger)$ with $\Omega = \sqrt{\kappa} |\bar{a}_\mathrm{in}|$. We seek the transition rates $\Gamma_{n \rightarrow m \neq n}$ from the state $\ket{n}$ with $n$ phonons to the state $\ket{m}$ with $m$ phonons induced by the optical drive. We work to second order in $\Omega$ to obtain a Fermi Golden rule result \cite{GirvinRMP} valid for $\Omega \ll \kappa$ where cavity states with more than one photon can be neglected. This gives
\begin{align}
\Gamma_{n\rightarrow m\not=n} = &\frac{1}{\hbar^2 t} \int_0^t \!\! d\tau_1 \!\! \int_0^t \!\! d\tau_2  \bra{n} \bra{i} \hat{H}'_1(\tau_1) \ket{m} \bra{m} \hat{H}'_1 (\tau_2) \ket{i} \ket{n} \nonumber \\
= &\frac{\kappa \Omega^2}{t} \int_0^t \!\! d\tau_1 \!\! \int_0^t \!\! d\tau_2 \!\! \int_{-\infty}^{\tau_1} \!\!\!\!\! ds_1 \!\! \int_{-\infty}^{\tau_2} \!\!\!\!\! ds_2 \,  \nonumber \\
& e^{-(\kappa/2-i\tilde{\Delta})(\tau_1-s_1)} e^{-(\kappa/2+i\tilde{\Delta})(\tau_2-s_2)} \nonumber \\
&\times \bra{i} \des{\xi}(s_1) \cre{\xi}(s_2) \ket{i} \bra{n} e^{\hat{X}(\tau_1)} e^{-\hat{X}(s_1)} \ket{m} \nonumber \\
&\times \bra{m}  e^{\hat{X}(s_2)} e^{-\hat{X}(\tau_2)}  \ket{n}
\label{rates1}
\end{align}
where $\tilde{\Delta} = \Delta + g^2/\omega_M$, $|i\rangle$ is the vacuum state of the optical bath, and we used the solution to Eq.~(\ref{motion1}) in the absence of an optical drive, as derived in the appendix.

Using a resolution of unity we rewrite the matrix element $\bra{n} e^{\hat{X}(\tau_1)} e^{-\hat{X}(s_1)} \ket{m} = \sum_k \bra{n} e^{\hat{X}(\tau_1)} \ket{k} \bra{k} e^{-\hat{X}(s_1)} \ket{m}$. For large mechanical quality factors we only need to consider the free mechanical evolution $\bra{n} e^{\hat{X}(\tau_1)} \ket{k} = e^{i(n-k) \omega_M \tau_1} Z_{n,k}$ where we have evaluated the matrix elements to be $Z_{n,k} = (-1)^{(n-k+|n-k|)/2} e^{-\lambda^2/2} \lambda^{|n-k|}\sqrt{\frac{\textrm{min}(n,k)!}{\textrm{max}(n,k)!}} L^{(|n-k|)}_{\textrm{min}(n,k)}(\lambda^2)$ with the associated Laguerre polynomials $L^{(\alpha)}_{n}(x)$ \cite{Abramowitz1964} and the coupling strength $\lambda = g/\omega_M$. Finally, we obtain the rates
\begin{align}
\Gamma_{n\rightarrow m\not=n} &= \kappa \Omega^2 \left| \sum_{k=0}^\infty \frac{Z_{n,k} Z_{m,k}}{\kappa/2 - i [(n-k)\omega_M + \tilde{\Delta}]} \right|^2.
\label{rates}
\end{align}
In the resolved-sideband limit $\omega_M \gg \kappa$ only terms with $k=n-l\geq0$ contribute significantly for detunings $\tilde{\Delta} \approx -l \omega_M$
\begin{align}
\Gamma_{n\rightarrow m\not=n} &= \frac{\kappa \Omega^2 Z_{n,n-l}^2 Z_{n-l,m}^2}{(\kappa/2)^2 + (l \omega_M + \tilde{\Delta})^2}.
\label{rates2}
\end{align}

Eqs.~(\ref{rates}) and (\ref{rates2}) are our main result from which we obtain a clear physical picture how an incident photon is inelastically scattered off the cavity and changes the state of the mechanical oscillator from $n$ to $m$ phonons. In the resolved-sideband limit $\omega_M \gg \kappa$ and for a drive detuned by $\tilde{\Delta} = -l\omega_M$, the process of destroying $l$ phonons when creating the cavity photon is enhanced by the cavity susceptibility. The amplitude for this process is proportional to the matrix element $Z_{n,n-l} = \int dx \, \varphi_{n-l}^*(x-x_0) \varphi_{n}(x)$ where $\varphi_{m}(x)$ are the eigenfunctions of the simple harmonic oscillator and $x_0 = -2 x_\mathrm{ZPF} g/\omega_M$ is the displacement caused by a single photon. That means that the matrix element is given by an overlap integral between displaced harmonic oscillator wave functions in accordance with the Franck-Condon principle. As the photon leaves the cavity it induces a transition in the mechanical oscillator from $n-l$ to $m$ phonons. This process is not resonantly enhanced as the photon decays into the continuum of modes in free space. This is why its amplitude is just given by the matrix element $Z_{n-l,m}$ which is a function of the ratio $\lambda^2 = g^2/\omega_M^2$, i.e.~the strength of the optomechanical interaction $g$ relative to the frequency of the mechanical oscillator $\omega_M$, but does not depend on the drive detuning $\Delta$. The photon in the output field has an energy $\hbar \omega_L + (n-m)\hbar \omega_M$, i.e.~it carries away the energy of $n-m$ phonons. In the non-resolved sideband limit $\omega_M \ll \kappa$ processes with different intermediate phonon number $n-k \not= l$ contribute, and their amplitudes interfere according to Eq.~(\ref{rates}).

In Fig.~\ref{fig:cooling} we plot the steady-state phonon number $\bar{n}= \langle \cre{b} \des{b} \rangle$ as a function of detuning $\Delta = \omega_L-\omega_R$ for different coupling strengths $g/\omega_M$ (a) and sideband parameters $\omega_M/\kappa$ (b). In (a) we observe that for weak drive and in the resolved-sideband limit $\omega_M \gg \kappa$ several cooling resonances appear when the detuning matches $\tilde{\Delta} = -l \omega_M$ with $l$ integer. For smaller sideband parameters (b) we notice that resonances merge.

\emph{Validity of the rate equation approach.} To investigate the validity of the set of rate equations (\ref{rateeqn}) we solve numerically the quantum master equation
\begin{equation}
\dot{\varrho} = - \frac{i}{\hbar} \left[ \op{H}', \varrho \right] + \kappa \mathcal{D}[\des{a}]\varrho + \gamma (1+n_\mathrm{th}) \mathcal{D}[\des{b}]\varrho + \gamma n_\mathrm{th} \mathcal{D}[\cre{b}]\varrho
\label{master}
\end{equation}
with $\op{H}' = \op{H}'_0 + \op{H}'_1$ where $\mathcal{D}[\des{o}] \varrho = \des{o} \varrho \cre{o} - (\cre{o} \des{o} \varrho + \varrho \cre{o} \des{o})/2$ is the standard dissipator in Lindblad form and $\op{H}'_0$ is the Hamiltonian (\ref{Ham1}) in the frame rotating at the drive frequency $\omega_L$.

In Fig.~\ref{fig:cooling} we plot the master equation (\ref{master}) results alongside with those from the set of rate equations (\ref{rateeqn}). The agreement between the two is excellent. The small deviations at detunings $\Delta \approx 0$ between the rate equation and master equation results stem from the $n$-photon resonances at $\Delta = -ng^2/\omega_M$ \cite{Nunnenkamp2011}. In their vicinity off-diagonal elements of the density matrix cannot be neglected and the rate equation approach fails.

\begin{figure}
\centering
\includegraphics[width=0.49\columnwidth]{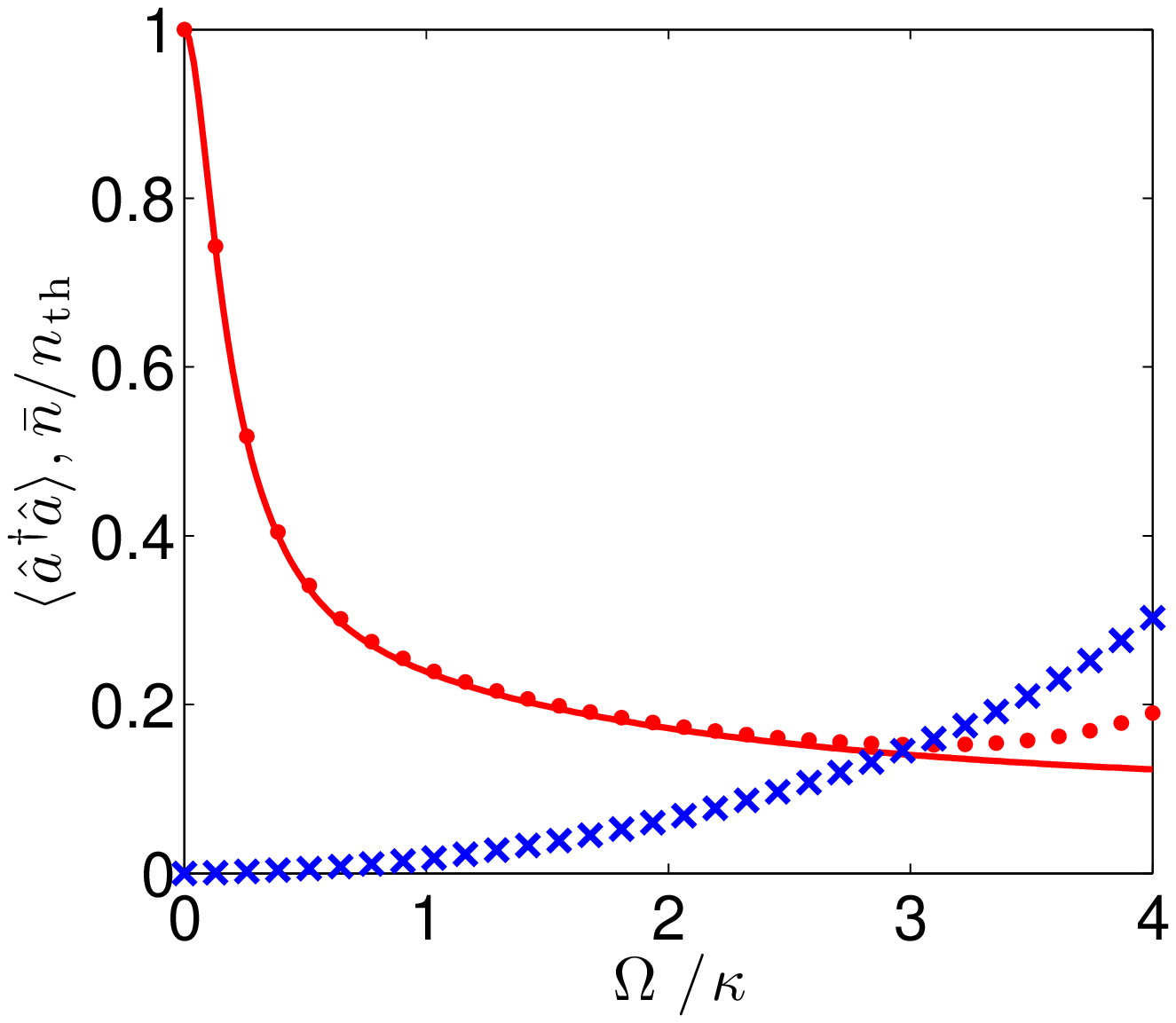}
\includegraphics[width=0.49\columnwidth]{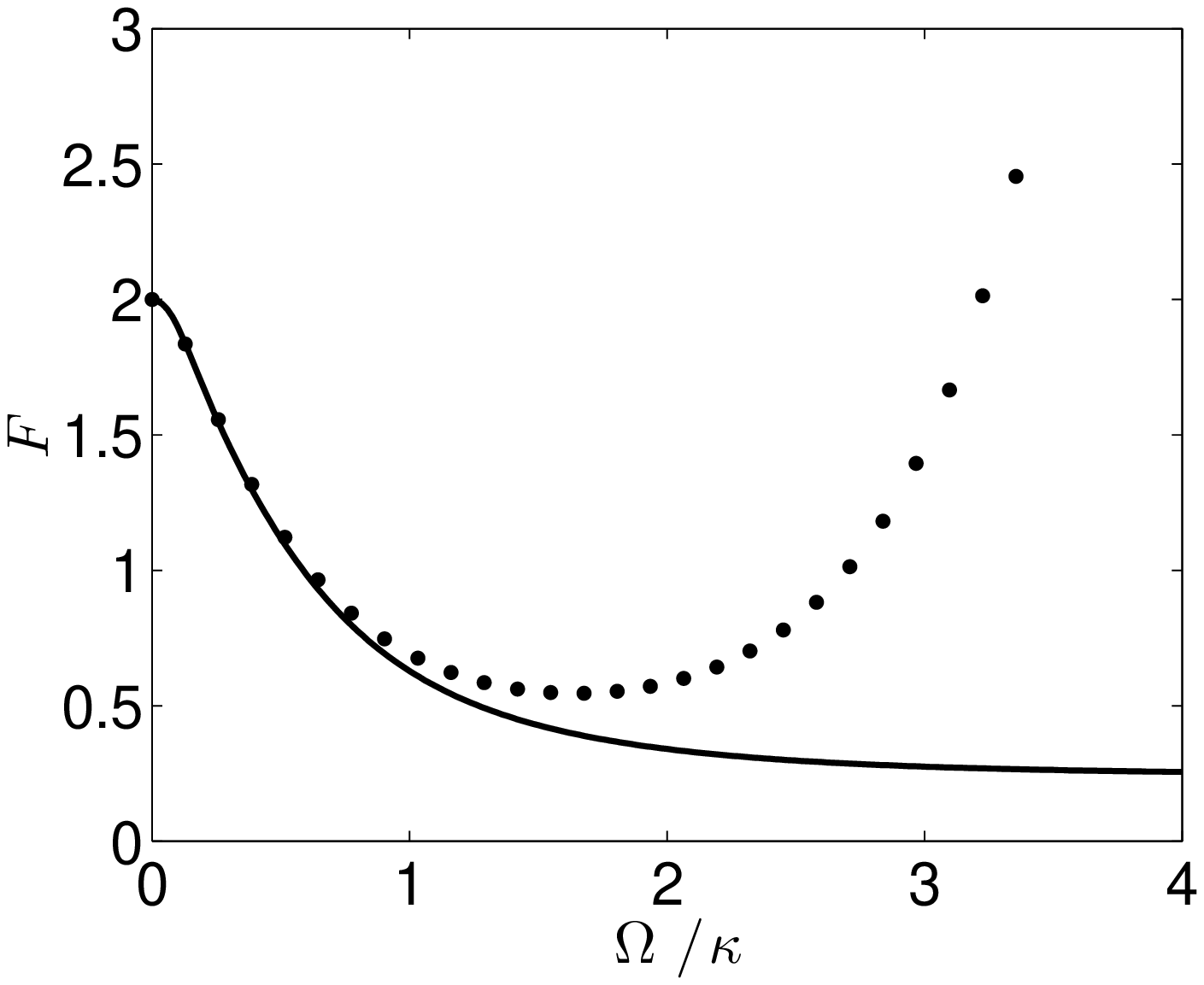}
\caption{(Color online) \emph{Validity of the rate equation approach and phonon anti-bunching.} (a) Mean photon number $\langle \cre{a} \des{a} \rangle$ (blue crosses) and mean phonon number $\bar{n} = \langle \cre{b} \des{b} \rangle$ (red solid) as well as (b) phonon number fluctuations $F = \langle \cre{b} \cre{b} \des{b} \des{b} \rangle / \langle \cre{b} \des{b} \rangle^2$ as a function of the drive strength $\Omega$ for a detuning $\tilde{\Delta}/\omega_M = -2$. Results from the set of rate equations (\ref{rateeqn}) are shown as lines and those from the quantum master equation (\ref{master}) as dots and crosses. The other parameters are $\omega_M/\kappa = 4$, $g/\omega_M = 0.5$, $\Omega/\kappa = 0.2$, and $\omega_M/\gamma = 1000$.}
\label{fig:cooling2}
\end{figure}

In Fig.~\ref{fig:cooling2} (a) we plot the mean photon and phonon number, $\langle \cre{a} \des{a} \rangle$ and $\bar{n} = \langle \cre{b} \des{b} \rangle$, as a function of the drive strength $\Omega$. Driving on the second red sideband, i.e.~$\tilde{\Delta}/\omega_M = -2$, we find that for small $\Omega/\kappa$ the phonon number $\bar{n}$ decreases quadratically in the drive strength $\Omega$ as expected. For large $\Omega/\kappa$ the set of rate equations (\ref{rateeqn}) predict a finite minimal phonon number determined by the optically induced rates (\ref{rates}). However, in Fig.~\ref{fig:cooling2} (a) we find that as the photon number $\langle \cre{a} \des{a}\rangle$ increases with stronger drive $\Omega$ and the master equation result (\ref{master}) starts to deviate from solution of the rate equations (\ref{rateeqn}).

\begin{figure}
\centering
\includegraphics[width=0.49\columnwidth]{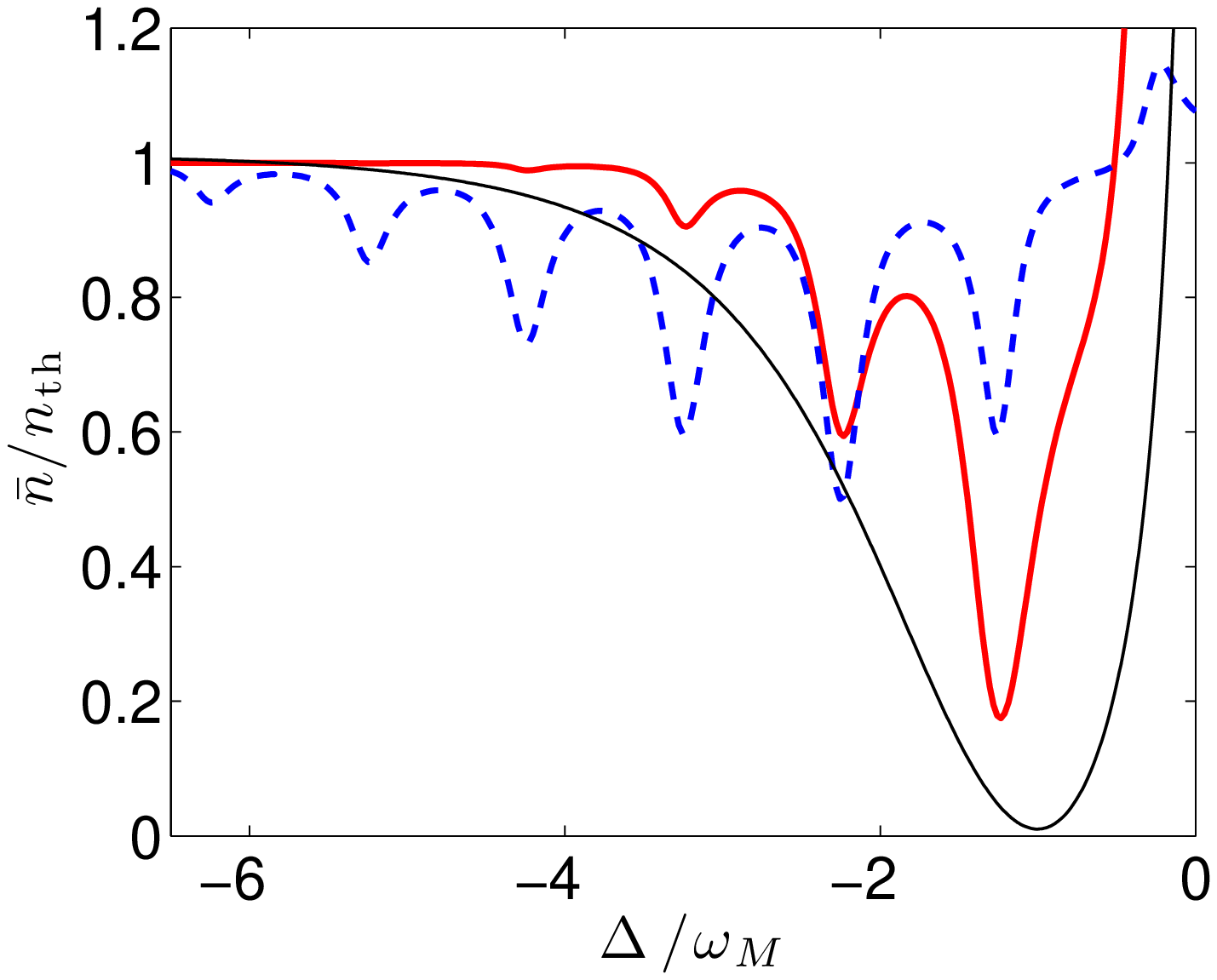}
\includegraphics[width=0.49\columnwidth]{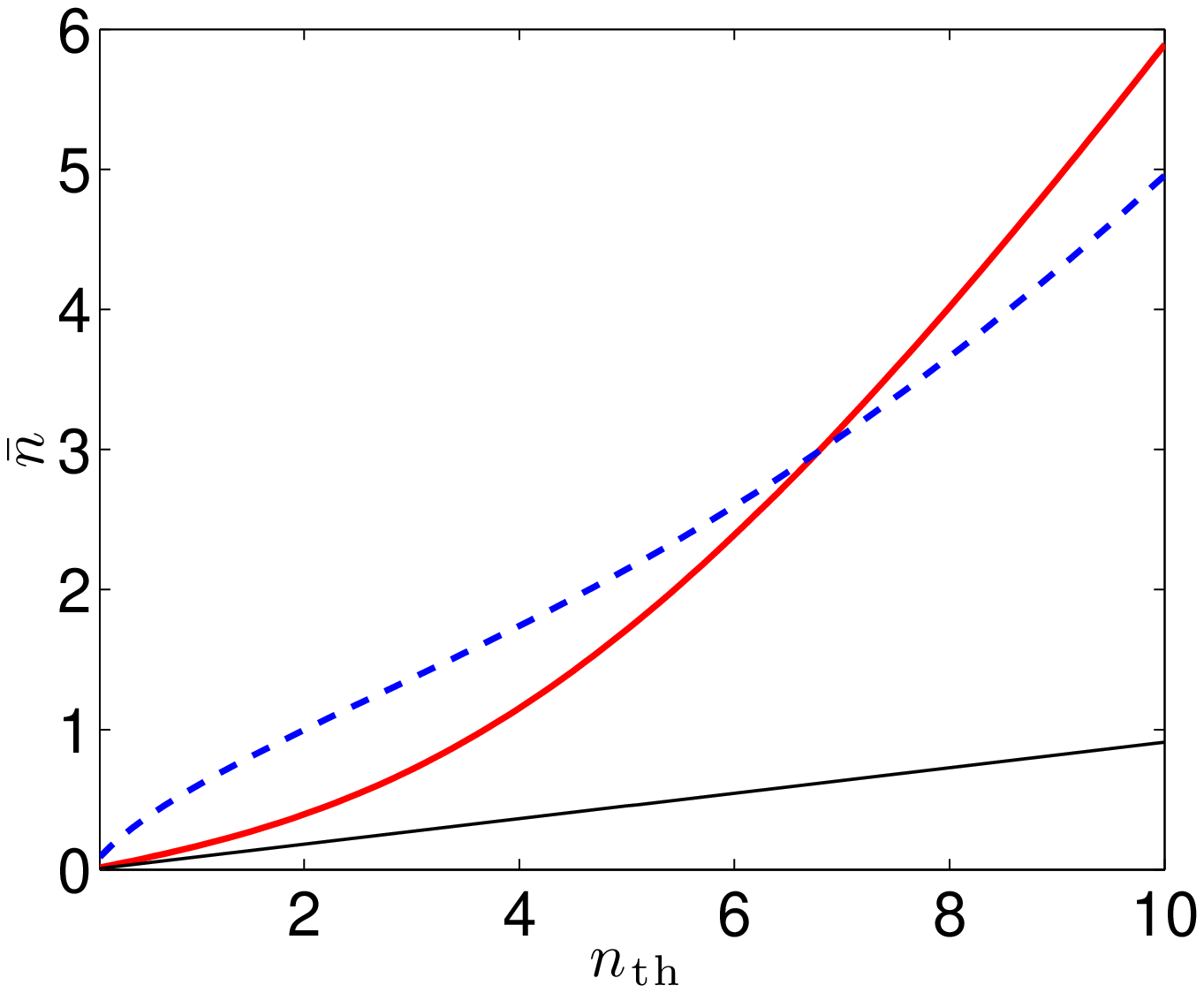}
\includegraphics[width=0.49\columnwidth]{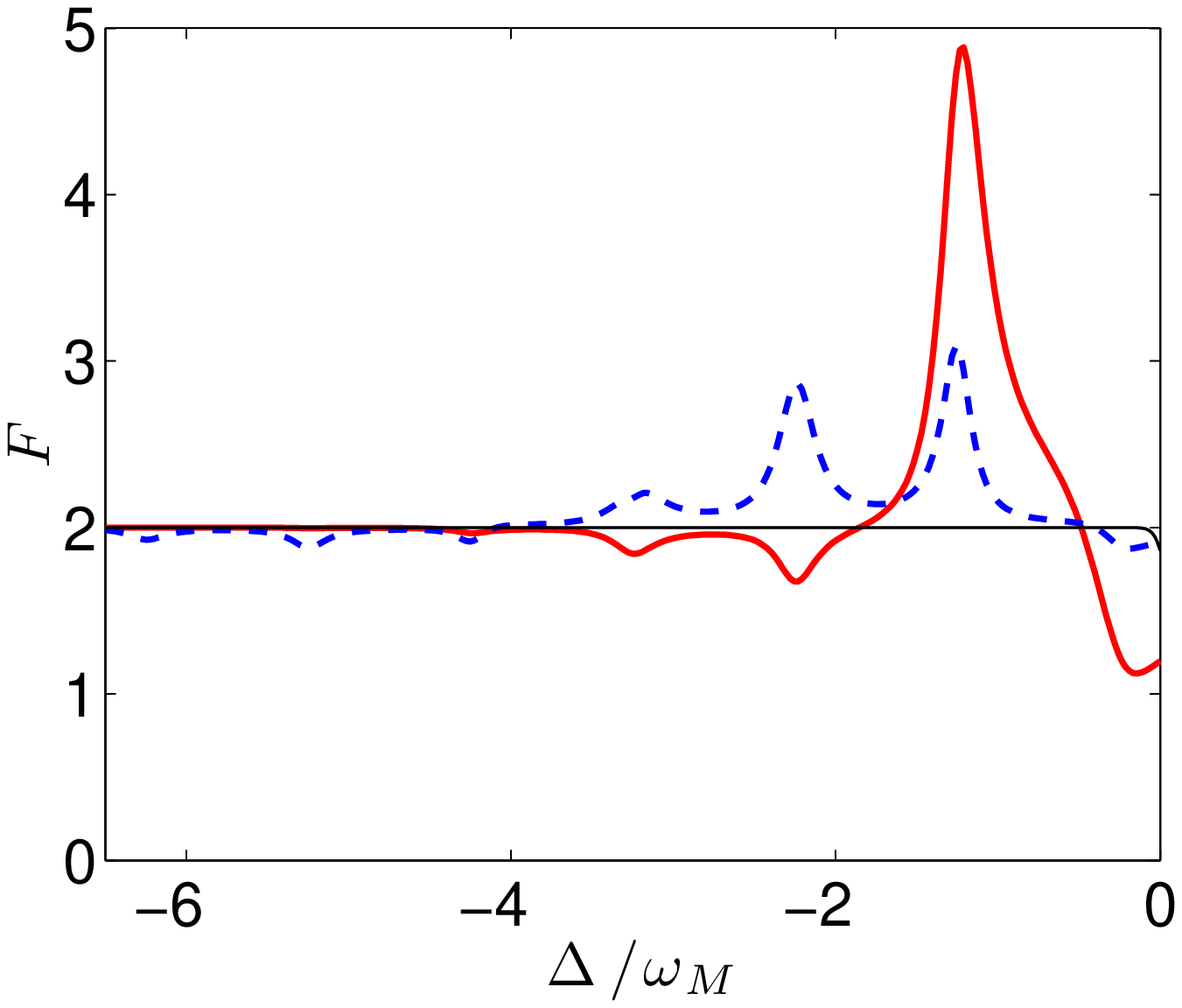}
\includegraphics[width=0.49\columnwidth]{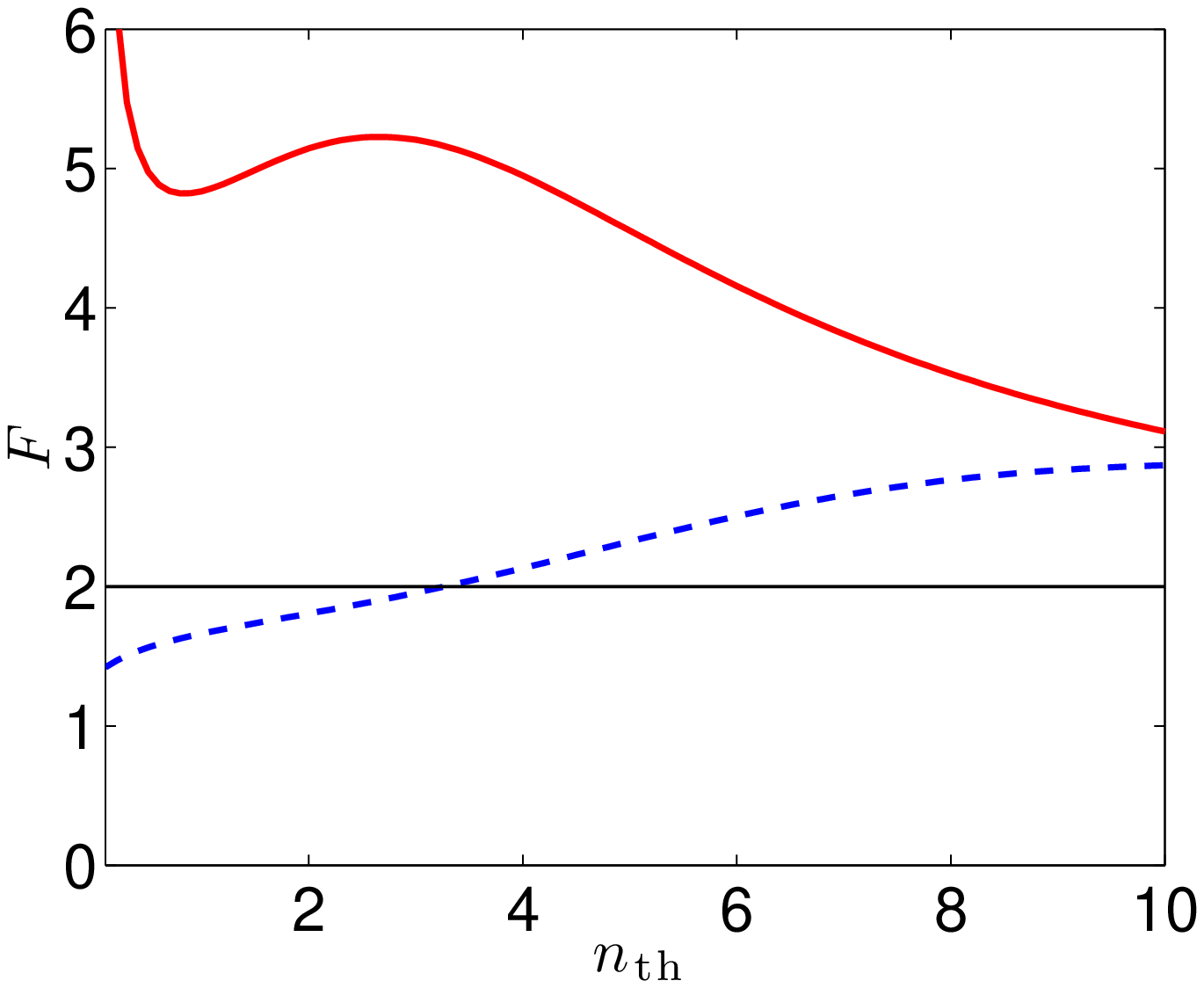}
\caption{(Color online) \emph{Non-thermal steady-states.} (Left) Phonon number $\bar{n} = \langle \cre{b} \des{b} \rangle$ (top) and phonon number fluctuations $F = \langle \cre{b} \cre{b} \des{b} \des{b} \rangle / \langle \cre{b} \des{b} \rangle^2$ (bottom) as a function of detuning $\Delta$ for $n_\mathrm{th} = 1$ (solid red) and $n_\mathrm{th} = 10$ (dashed blue). (Right) Phonon number $\bar{n}$ (top) and phonon number fluctuations $F$ (bottom) as a function of the thermal phonon number $n_\mathrm{th}$ for $\tilde{\Delta} = -\omega_M$ (solid red) and $\tilde{\Delta} = -2\omega_M$ (dashed blue). The other parameters are $\omega_M/\kappa = 4$, $g/\omega_M = 0.5$, $\Omega/\kappa = 0.2$, and $\omega_M/\gamma = 1000$. Thin black lines show results of the linear model (\ref{down}) and (\ref{up}) for $\Omega\lambda/\kappa = 0.2$.}
\label{fig:cooling3}
\end{figure}

\emph{Non-thermal steady states.} The rate equations (\ref{rateeqn}) in general do not obey detailed balance and so their steady state will not be a thermal state. In Fig.~\ref{fig:cooling3} we plot the mean phonon number $\bar{n}= \langle \cre{b} \des{b} \rangle$ and the number fluctuations $F = \langle \cre{b} \cre{b} \des{b} \des{b} \rangle / \langle \cre{b} \des{b} \rangle^2$ as a function of detuning $\Delta$. We note that the cooling power $\bar{n}/n_\mathrm{th}$ of the various cooling resonances depends on the thermal phonon number $n_\mathrm{th}$. This is a consequence of the nonlinear dependence of the rates (\ref{rates}) on phonon number. Close to the cooling resonances the mechanical state clearly deviates from a thermal state whose number fluctuations are given by $F = 2$. We find both reduced $F<2$ and enhanced number fluctuations $F>2$. In Fig.~\ref{fig:cooling3} we also plot the mean phonon number $\bar{n}$ and number fluctuations $F$ as a function of the thermal phonon number $n_\mathrm{th}$ for $\tilde{\Delta} = -\omega_M$ and $\tilde{\Delta} = -2\omega_M$. The mean phonon number $\bar{n}$ is a nonlinear function of the thermal phonon number $n_\mathrm{th}$ and the fluctuations $F$ can  change from $F<2$ to $F>2$ as a function of $n_\mathrm{th}$. In Fig.~\ref{fig:cooling2} (b) we find that even phonon anti-bunching, i.e.~$F<1$, can occur.

To understand this behavior let us look at the properties of the rate equations in more detail. For example, the resonant one-phonon cooling $\Gamma_{n \rightarrow n-1}$ and amplification $\Gamma_{n \rightarrow n+1}$ rates in the resolved-sideband limit $\omega_M \gg \kappa$ (\ref{rates2}) read
\begin{equation}
\frac{\Gamma_{n \rightarrow n-1}}{n\Gamma_{1\rightarrow 0}} = \frac{[L^{(1)}_{n-1}(\lambda^2) L^{(0)}_{n-1}(\lambda^2)]^2}{n^2}
\label{ratedown}
\end{equation}
and
\begin{equation}
\frac{\Gamma_{n \rightarrow n+1}}{(n+1) \Gamma_{1\rightarrow 0}} = \frac{[L^{(1)}_n(\lambda^2) L^{(0)}_{n+1}(\lambda^2)]^2}{(n+1)^2}.
\label{rateup}
\end{equation}

In the special case of weak coupling $\lambda \ll 1$ where one-phonon processes are most important, since higher-order processes are suppressed by a larger power of $\lambda$, we obtain
\begin{equation}
\Gamma_{n\rightarrow n-1} = \frac{\kappa \Omega^2 \lambda^2 n}{(\kappa/2)^2 + (\tilde{\Delta} + \omega_M)^2}
\label{down}
\end{equation}
and
\begin{equation}
\Gamma_{n\rightarrow n+1} = \frac{\kappa \Omega^2 \lambda^2 (n+1)}{(\kappa/2)^2 + (\tilde{\Delta} - \omega_M)^2}.
\label{up}
\end{equation}
That means we recover the standard cooling theory \cite{Wilson-Rae2007, Marquardt2007} where the rates are linear in $n$ and $n+1$, respectively, i.e.~one can write them as $\Gamma_{n \rightarrow n-1} = n \Gamma_\downarrow$ and $\Gamma_{n \rightarrow n+1} = (n+1) \Gamma_\uparrow$. In this case $\Gamma_\downarrow$ and $\Gamma_\uparrow$ simply renormalize the thermal mean phonon number $\bar{n} = (\gamma n_\mathrm{th} + \Gamma_\uparrow)/(\gamma + \Gamma_\downarrow - \Gamma_\uparrow)$. Thus, the steady state is a thermal state with $F=2$ for all detunings $\Delta$ and the mean phonon number $\bar{n}$ is linear in the thermal phonon number $n_\mathrm{th}$. In Fig.~\ref{fig:cooling3} we plot the mean phonon number $\bar{n}$ and the phonon number fluctuations $F$ for the weak-coupling limit $\lambda \ll 1$. We find a single cooling resonance at $\tilde{\Delta} = -\omega_M$ and $F=2$, indicative of a thermal state.

In general, the normalized rates (\ref{ratedown}) and (\ref{rateup}) differ from unity and multiple-phonon processes $\Gamma_{n \rightarrow m}$ with $|n-m|>1$ are important. If only a few phonons remain, further insight can be obtained. For example, in Fig.~\ref{fig:cooling2} (b) for $\Omega/\kappa \approx 2$, only the states with zero, one and two phonons are important. The two-phonon cooling process $\Gamma_{2 \rightarrow 0}$ reduces the occupation in the two-phonon state leading to phonon anti-bunching $F<1$ similar to the case studied in Ref.~\cite{Nunnenkamp2010}. In contrast, in Fig.~\ref{fig:cooling3} for $n_\textrm{th} = 1$ and $\tilde{\Delta} = -\omega_M$, we have $\Gamma_{2 \rightarrow 1}/(2 \Gamma_{1\rightarrow 0}) < 1$, i.e.~relative to a thermal state the one-phonon state is depleted faster than the two-phonon state which results in $F>2$.

\emph{Detection.} The multiple-phonon cooling and amplification processes lead to multiple mechanical sidebands in the optical output spectrum \cite{Nunnenkamp2011}. This is not a proof of non-thermal states by itself, as it can also occur e.g.~in the case of large mechanical amplitude motion at weak optomechanical coupling \cite{Hertzberg2010}. The mean phonon number $\bar{n}$ and the phonon number fluctuations $F$ can be obtained from the Wigner function of the mechanical oscillator. Schemes to reconstruct the Wigner function experimentally rely on back-action evasion \cite{Clerk2008}, coupling to a two-level system \cite{Singh2010, Tufarelli2011}, or the time-dependence induced by short optical pulses \cite{Vanner2011}.

\emph{Acknowledgements.} We are grateful for support by the Swiss National Science Foundation through the NCCR Quantum Science and Technology (AN), from the NSF under Grant No.~DMR-1004406 (AN and SMG) and DMR-0653377 (SMG), from the Research Council of Norway under Grant No.~191576/V30 (KB) and The Danish Council for Independent Research under the Sapere Aude program (KB). Part of the calculations were performed with the Quantum Optics Toolbox \cite{Tan1999}.

\emph{Appendix.} To calculate the optically induced transition rates between different phonon Fock states (\ref{rates1}), we need an expression for the optical field in the absence of an optical drive. The formal solution to Eq.~\eqref{motion1} is 
\begin{equation}
\hat{a}(t) = \sqrt{\kappa} \int_{-\infty}^t \!\!\!\!\!\!  d \tau \hat{K}(t,\tau) \hat{a}_\mathrm{in}(\tau) 
\end{equation}
where $\hat{K}(t,\tau) = e^{-\left(\kappa/2 - i \Delta\right) (t - \tau)} {\cal T} \left[e^{-i g \int_{\tau}^t ds (\hat{b}(s) + \hat{b}^\dagger(s))}\right]$ and ${\cal T}$ is the time ordering operator. In the case of no optical drive, $\bar{a}_\mathrm{in} = 0$, the operator identity $\hat{a}^\dagger \hat{a} = 0$ holds, such that $\hat{b}(t') = e^{-i \omega_M (t' - t)} \hat{b}(t)$ for times $|t' - t| \ll (\gamma n_\mathrm{th})^{-1}$. This means that for times $(t - \tau) \ll (\gamma n_\mathrm{th})^{-1}$, we can express the time ordered exponential above as $e^{i g^2(t - \tau)/\omega_M} e^{\hat{X}(t)} e^{-\hat{X}(\tau)}$ where $\hat{X}(t) = g[\hat{b}(t) - \hat{b}^\dagger(t)]/\omega_M$. This follows from using the standard commutation relations for the bosonic operator $\hat{b}$. Thus, in the limit $\gamma n_\mathrm{th} \ll \kappa$, we find
\begin{equation}
\hat{a}(t) = \sqrt{\kappa} \int_{-\infty}^t \!\!\!\!\!\! d \tau  e^{-\left(\kappa/2 - i \tilde{\Delta}\right) (t - \tau)} e^{\hat{X}(t)} e^{-\hat{X}(\tau)} \hat{\xi}(\tau)
\end{equation}
where $\tilde{\Delta} = \Delta + g^2/\omega_M$. This expression can also be derived using the polaron transform \cite{Nunnenkamp2011}.

\end{document}